\documentclass[conference]{IEEEtran}

\usepackage[letterpaper, left=1in, right=1in, bottom=1in, top=0.75in]{geometry}

\ifCLASSINFOpdf
   \usepackage[pdftex]{graphicx}
   \DeclareGraphicsExtensions{.eps,.jpg,.png}
\else

\fi

\usepackage{amsmath}
\usepackage{amssymb}
\usepackage{mathtools}
\usepackage{booktabs}

\begin{document}

\title{Renewal-Theoretic Packet Collision Modeling under Long-Tailed Heterogeneous Traffic}

\author{\IEEEauthorblockN{Aamir~Mahmood and Mikael~Gidlund}
\IEEEauthorblockA{Department of Information Systems and Technology\\
Mid Sweden University, Sweden\\
Email: firstname.lastname@miun.se}}

\maketitle

\begin{abstract} 

Internet-of-things (IoT), with the vision of billions of connected devices, 
is bringing a massively heterogeneous character to wireless connectivity in 
unlicensed bands. The heterogeneity in medium access parameters, transmit 
power and activity levels among the coexisting networks leads to detrimental 
cross-technology interference. The stochastic traffic distributions, shaped 
under CSMA/CA rules, of an interfering network and channel fading makes it 
challenging to model and analyze the performance of an interfered network. In 
this paper, to study the temporal interaction between the traffic 
distributions of  two coexisting networks, we develop a renewal-theoretic 
packet collision model and derive a generic collision-time distribution (CTD) 
function of an interfered system. The CTD function holds for any busy- and 
idle-time distributions of the coexisting traffic. As the earlier studies 
suggest a long-tailed idle-time statistics in real environments, the 
developed model only requires the Laplace transform of long-tailed 
distributions to find the CTD. Furthermore, we present a packet error rate 
(PER) model under the proposed CTD and multipath fading of the interfering 
signals. Using this model, a computationally efficient PER approximation for 
interference-limited case is developed to analyze the performance of an 
interfered link.

%
%Furthermore, we develop a packet 
%error rate (PER) model under the proposed CTD and fading of the interfering signals, and 
%develop its computationally efficient approximations. 

%Other than performance 
%modeling, the proposed packet collision model finds its utility in 
%simulations for coexistence analysis and RF energy harvesting 
%studies; for instance, to find energy distribution under stochastic 
%activation of the RF source and the harvesting device. 

\end{abstract}

\section{Introduction}

In typical office, home and industrial settings, simultaneous presence of 
heterogeneous wireless technologies is becoming certain; now that we are \emph
{on the pulse of the networked society} \cite{ericsson2015}. For instance, 
we use WLAN for the Internet, and low-power Bluetooth- and IEEE 802.15.4-
based HVAC and industrial control systems. The coexistence of these 
technologies affects their performance in three domains: frequency, time, and 
space. On a certain frequency channel, interference in temporal domain is 
dictated by the traffic parameters whereas the spatial interference depends 
on the transmission power and location of the interferer, and multipath 
fading. Be it temporal or spatial domain, modeling the heterogeneous 
coexistence to its exactness is quite complex although desired for 
performance evaluation and enhancement especially in interference prone 
low-power networks.

%Regardless of the considered interference domain, low-power networks 
%are at the losing end in coexistence scenarios. 

 Starting with temporal overlap between two coexisting networks, the 
collision-time of interfered packets together with the SINR determines the eventual 
packet error rate (PER). The collision-time is defined by the traffic 
parameters of the coexisting networks: that is, distributions of the packet 
length and idle-time. In a realistic environment, while modeling the 
collisions with a multi-terminal system like WLAN, the compound traffic 
observed by an interfered link has to be considered. In many measurement 
studies \cite{LucaStab}\cite{geirhofer2006measurement}, it is shown that the 
WLAN traffic shaped under CSMA/CA protocol follows long-tailed idle-time 
statistics such as hyperexponential or hyper-Erlang distribution. This is 
where the deterministic models (e.g., \cite{Shin}) for constant packet 
inter-arrivals fail to encompass the real traffic characteristics. A 
collision-time distribution for interfered packets of constant length in the 
presence of arbitrary idle-time (busy-time) statistics of the interference is 
developed in \cite{pcmAamir}. However, this distribution is derived for 
constant, and exponential and gamma distributions of the idle-time. 

In this paper, we derive a theoretical collision time distribution (CTD)   
which holds for any idle-time distribution with known Laplace transform. 
Thus, CTD can easily be evaluated for mixture distributions such as 
hyperexponential and hyper-Erlang. Using the alternating renewal process 
representation of the WLAN traffic from \cite{pcmAamir}, the 
collision-time of an interfered packet depends on the initial observed state 
of the WLAN traffic (i.e., idle or busy) as well as the residual life of that 
state. Then the CTD in each state is the distribution of the random sum of 
the busy-times  
encountered in the interfered packet length. While the random sum is 
weighted by the distribution of the number of renewals (idle-times) 
observed during the interfered packet. In particular for exponential 
distributed interfered packet length, we develop the distributions of the 
number of renewals for each initial observed state: which are generic and can 
easily be evaluated for any idle-time distribution. We validate the 
theoretical derived CTD with the simulation results, showing a perfect 
match. We also give the collision-time distributions for hyperexponential 
idle-times with parameters, reported in \cite{LucaStab}, fitted to the measurements from a real 
heterogeneous environment.       

The developed CTD and the distribution of the number of renewals under realistic 
coexisting traffic can be utilized in a number of ways e.g.,
\begin{itemize}
	\item Link quality analysis as studied further in this paper.
	\item Simulating the performance of a transmission scheme, 
packet length optimization and dimensioning the spectrum sensing algorithms.
	\item To find the distribution of the harvested energy based on the temporal overlap 
of RF power source and the harvesting device \cite{lee2013opportunistic}.
\end{itemize}

With the temporal interaction of an interfered link fully captured by 
collision-time distribution, we develop its PER model which, contrary to 
\cite{Shin}\cite{pcmAamir}, also incorporates the effect of multipath fading 
of the interfering signals during the collision time. Specifically, we 
consider the PER analysis of an interfered link operating in a relatively 
static environment, and in the presence of multiple interferers of identical 
powers undergoing Rayleigh fading. For the considered case, we develop two 
PER approximations for transmission schemes with bit error rate (BER) 
in the form of Gaussian $Q$-function. We evaluate the accuracy of each approximation and 
discuss how to combine them to evaluate the PER accurately and in 
computationally efficient manner.

%Specifically, the PER model assumes for an interfered link subject to constant channel gain,   
%
%Furthermore, we develop a PER model which besides CTD model incorporates the 
%Rayleigh block-fading of the interference. Assuming the constant channel gain 
%of an interference-limited link, we a develop a computationally efficient and accurate 
%approximation of the PER.

The rest of the paper is organized as follows. Section \ref{sec:pcmodel} 
develops a collision-time distribution function based on renewal-theoretic 
packet collision modeling. Section \ref{sec:CollisionTimeAnalysis} finds the 
distributions for interference \emph{on} time and the number of renewals. 
Section \ref{sec:PER} presents the PER model and develops its approximations. 
Section \ref{sec:Conclusions} draws the concluding remarks.
\section{System Model}
\label{sec:pcmodel}

\begin{figure}[t]
	\centering
		\includegraphics[width=0.60\linewidth]{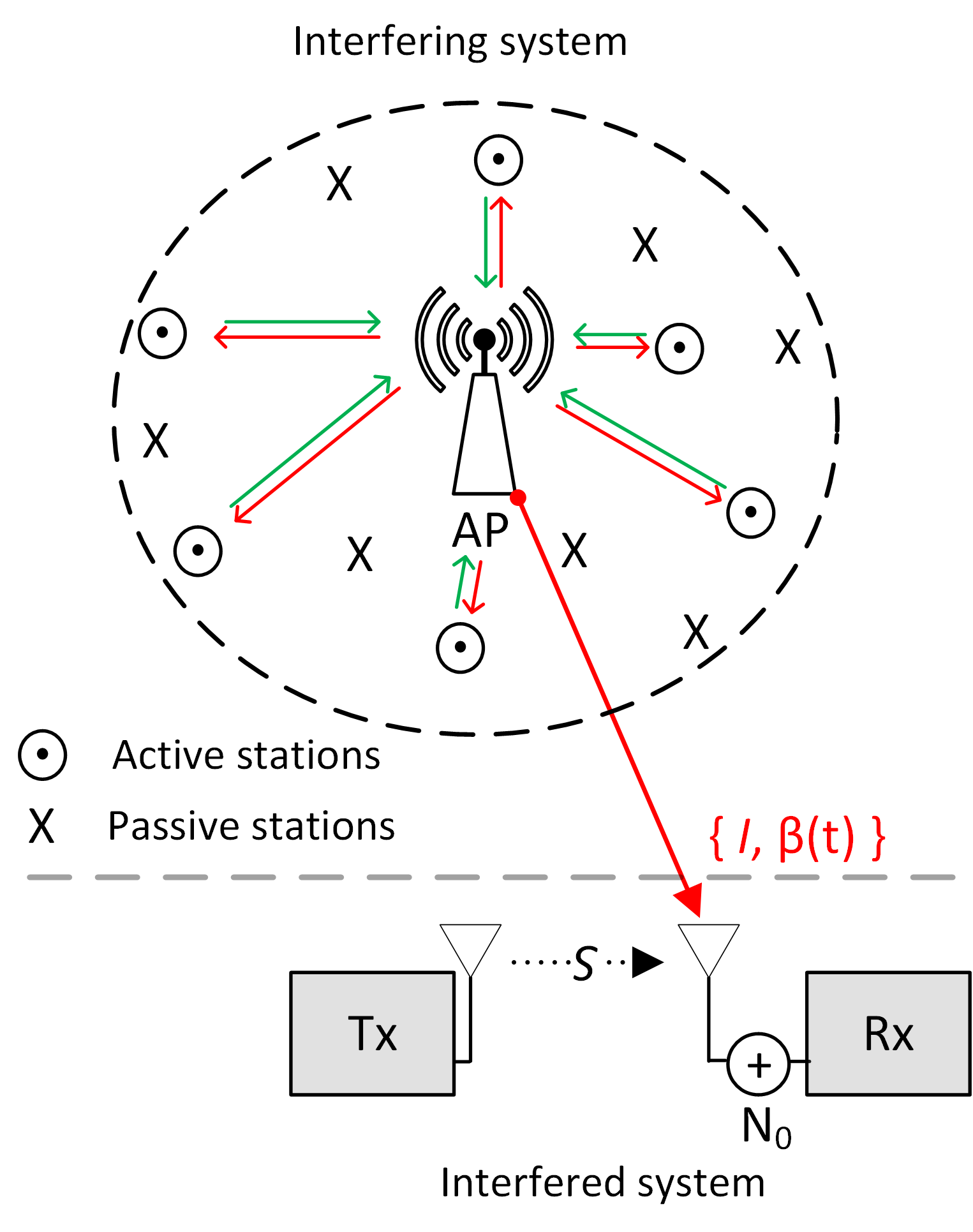}
	\caption{System model for heterogeneous coexistence}
	\label{fig:SysModel}
\end{figure}

We consider a low-power wireless sensor system, where each sensor link 
between a transmitter and receiver pair, is subjected to interference from 
coexisting WLAN system as shown in Fig.~\ref{fig:SysModel}. In the WLAN 
system, at any time instant, there can be a random number of associated 
stations however under CSMA/CA medium access (ideally) only one station is in 
transmit (receive) state to (from) the access point (AP). The composite 
traffic arrival process, shaped by CSMA/CA rules and with or without any 
perturbations in the arrival process due to collisions or interference, 
observed by the sensor link is denoted as $\beta(t)$ in 
Fig.~\ref{fig:SysModel}.

%When the system is
%interference-limited, the effect of thermal noise may be ignored
%[7].
%
%To further simplify
%the analysis, and since we are concerned with systems which
%are clearly interference limited, the effect of thermal noise is
%ignored in this paper.
%

\subsection{Packet Collision Model for an Interfered Sensor Link}
Now we develop a packet collision model for the sensor link under the effect of packet arrival process of the WLAN system.
\begin{figure}[t]
	\centering
		\includegraphics[width=0.70\linewidth]{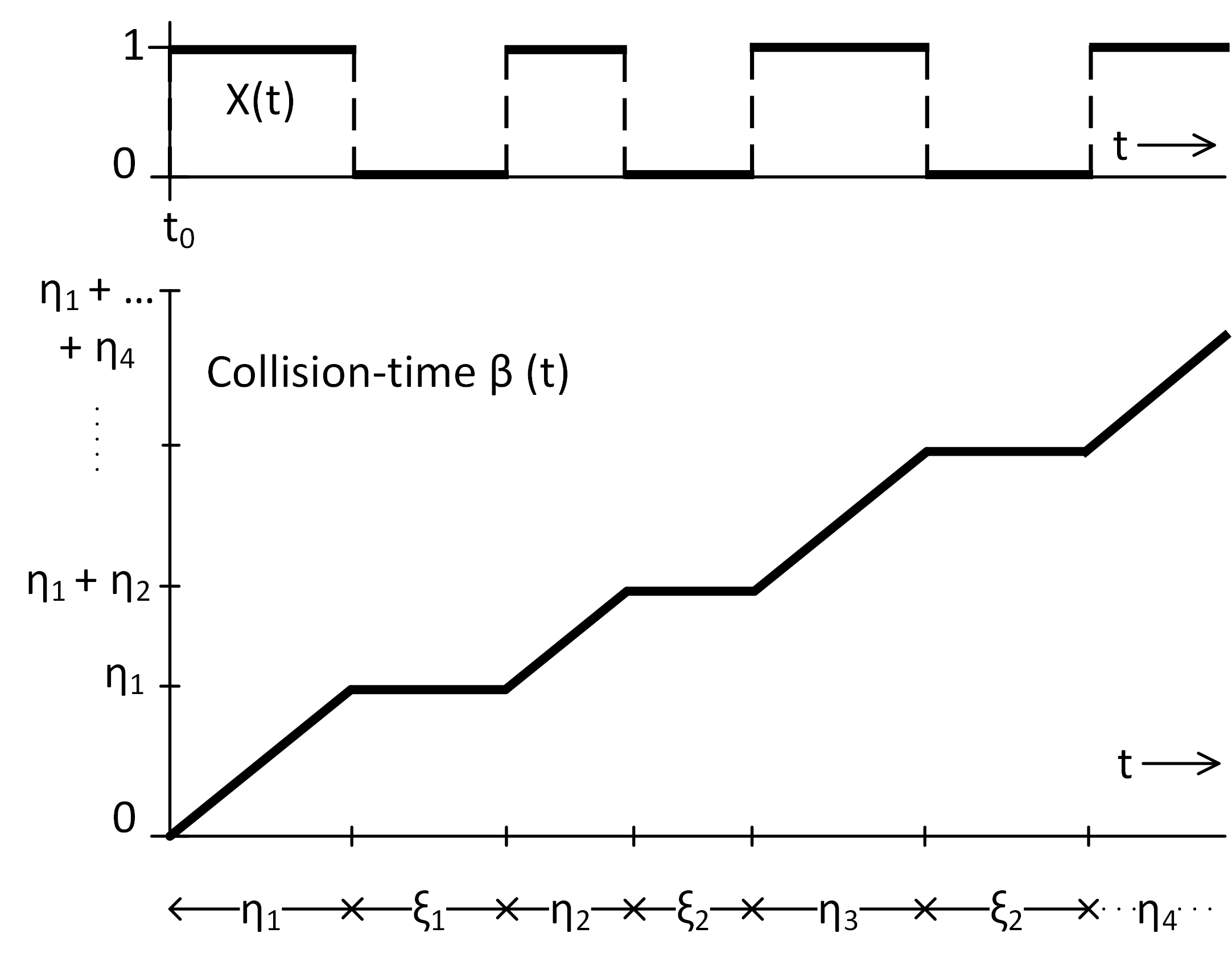}
	\caption{Sample functions of WLAN traffic and collision time processes}
	\label{fig:Sample_functions}
\end{figure}

Consider an alternating renewal process representing WLAN packet arrivals. At 
any time instant, the process is in one of the two states: busy (\emph{on}) or idle 
(\emph{off}) (see Fig.~\ref{fig:Sample_functions}). Denote the state of the process at time $t \geq 0$ by 
$\chi(t)$, and let $\chi(t) = 1$ if the process is \emph{on} and $\chi(t) = 0$
 if it is \emph{off}. The time evolution of the process is then described by 
the two state stochastic process $\{\chi(t), t\geq 0\}$. The total \emph{on} time in which  
the process spends in state $\chi(t) = 1$ during the time interval 
$(t_0, t)$ is a stochastic process, denoted as $\beta(t)$, and it is the 
potential collision time for a sensor link. Mathematically, $\beta(t)$ 
in terms of $\chi(t)$ is defined as
\begin{equation*}
\beta\left(t\right) = \int\limits_{t_0}^{t} \chi\left(x\right)\textnormal{d}x
\label{eq:Operation_time_integral}
\end{equation*}
The complement of $\beta(t)$ is the total \emph{off} time, which is the time
the process is spends in the state $\chi(t) = 0$ during $(t_0,t)$, and follows 
$\alpha\left(t\right) = t -  \beta\left(t\right)$.

Let variables $\eta_i(\xi_i)$ denote the time spent in state 
\emph{on}(\emph{off}) during the $i$th visit to that state. We assume 
that all $\eta_i(\xi_i)$ are independent and identically distributed as 
$\eta(\xi)$ according to 
$H\left(x\right) = \Pr\{\eta \leq x\}\left(G\left(x\right) = \Pr\{\xi \leq x
\}\right)$, and both $\eta$ and $\xi$ are continuous random variables with mean 
$\bar {\eta}$ and $\bar{\xi}$. The distribution of the sum $\eta_1 + \eta_2 +
\cdots + \eta_n$ is the $n$-fold convolution of $H(x)$, i.e. 
\begin{equation*}
H_n\left(x\right) = \Pr \Big\{\sum\limits_{i=1}^{n}\eta_{i} \leq x \Big\}
\end{equation*} 
where $H_0\left(x\right) = 1$. An analogous definition holds for 
$G_n\left(x\right)$, being the $n$-fold convolution of $G(x)$. A symbolic 
representation of the functions $\chi(t)$ and $\beta(t)$ is shown in 
Fig.~\ref{fig:Sample_functions}    

Assuming the process enters the state \emph{on} at time $t_0$, the cumulative 
distribution function (CDF) of the \emph{on} time, $\beta\left(t\right)$, is 
given by Tak{\'a}cs~\cite{Takacs1957}
\begin{equation}
\Pr \left\{\beta\left(t\right)\leq x \right\} = \sum_{n=0}^{\infty} {H}_{n
} \left(x \right) \Pr\{N\left(t-x\right) = n \}
\label{eq:Tackas_distribution}
\end{equation}
where $\Pr\{N\left(t-x\right) = n \}$, is the probability mass function (PMF) 
of the number of renewals (or the idle-times) during the 
time interval $(t_0, t-x)$ with $x$- the collision time. 
From \cite{cox1970renewal}, the PMF is related to $G_n(x)$ as 
\begin{equation}
\Pr\{N\left(t-x\right) = n \} = G_n\left(t-x\right) - G_{n+1}\left(t-x\right)
\label{eq:Nt}
\end{equation} 

\begin{figure}
	\centering
		\includegraphics[width=1\linewidth]{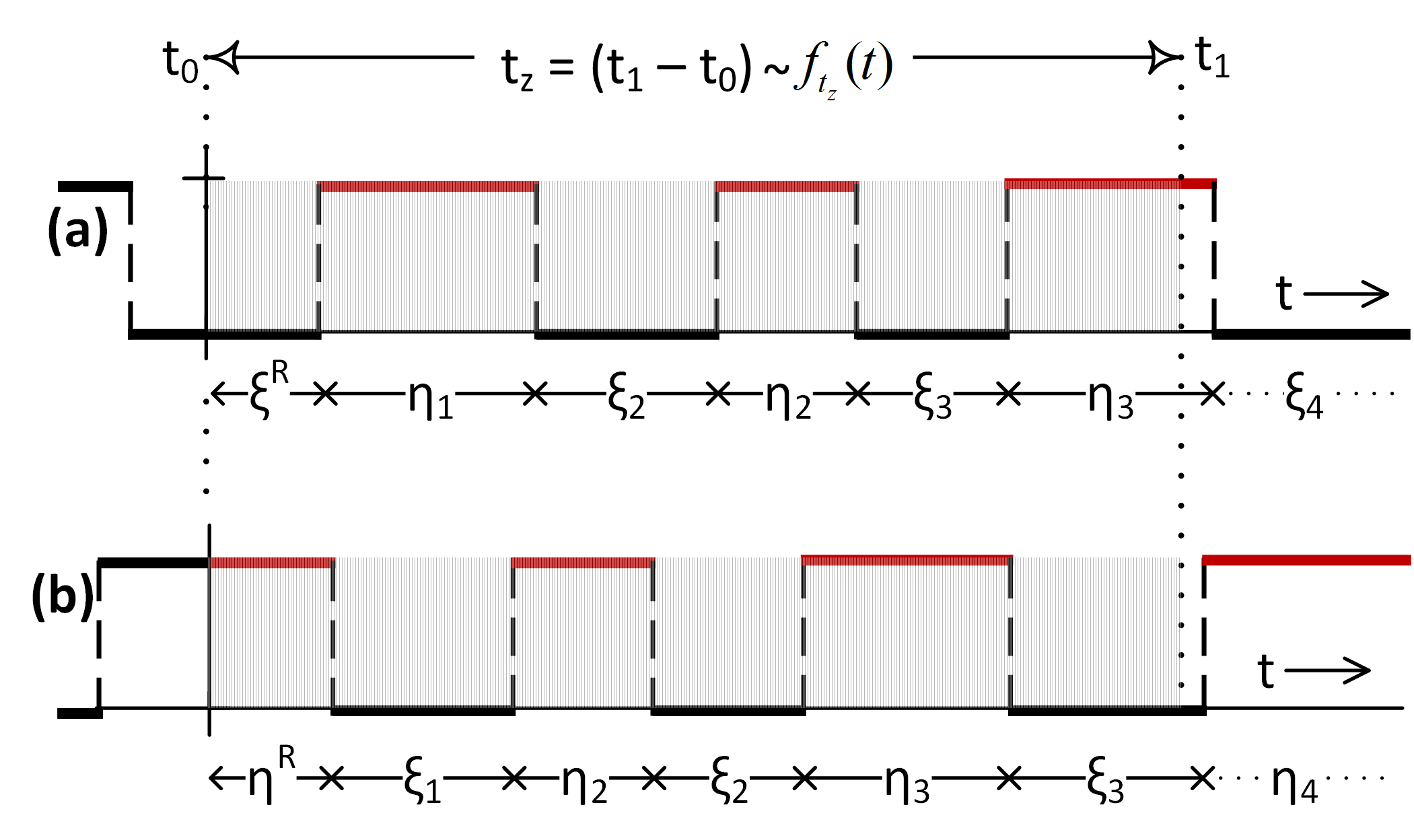}
	\caption{State of the alternating renewal process at observation instant: (a) there is no WLAN packet at the observation instant $t_0$ (that is, at the start of a transmission over the sensor link), (b) there is an ongoing WLAN packet transmission at $t_0$. The shaded area represents length of the interfered packet which follows a distribution.}
	\label{fig:CollisionModel}
\end{figure}

From an interfered system's perspective, the process $\chi\left(t\right)$ however 
can be in an arbitrary state at time instant $t_0$ (see Fig.~\ref{fig:CollisionModel}). As a result, 
at $t_0$ the $\chi\left(t\right)$ can be in either state \emph{on} or \emph{off} as 
shown in Fig.~\ref{fig:CollisionModel}(a) and Fig.~\ref{fig:CollisionModel}(b). Define $\eta^R$ and $\xi^R$ be the residual time of 
$\xi$ and $\eta$ with distribution function $H^R=\Pr\{\eta^R \leq x\}$ and $G^R=
\Pr\{\xi^R \leq x\}$ respectively. Then, the distribution of the sum 
$\eta^R + \eta_2 + \cdots + \eta_n$ is     
\begin{align}
H_n^{R}\left(x\right) & = H^R(x) \ast H_n(x) \nonumber \\
											& = \Pr \Big\{\eta^R + \sum\limits_{i=2}^{n}\eta_{i} \leq x \Big\}
\label{eq:H_n_R}
\end{align} 
%\begin{align*}
%G_n^{R}\left(x\right) & = \Pr \Big\{\xi^R + \sum\limits_{i=2}^{n}\xi_{i} \leq x \Big\}, G_0\left(x\right)=1 \\
										  %& = G^R(x) \ast G_n(x)
%\end{align*} 
with $H_0^R\left(x\right) = 1$ and $H_1^R\left(x\right) = H^R\left(x\right)$. 
An analogous definition holds for $G_n^R\left(x\right)$, being the 
convolution of $G_n(x)$ with $G^R(x)$. 

Now consider the packet length, $t_z$, of the interfered system is a random 
variable, independent of the random variables $\eta$ and $\xi$, with 
probability density function (pdf) $f_{t_z}(t)$. If $\chi\left(t_0\right)=0$, the CDF of \emph{on} time (i.e., the collision-time 
distribution of the sensor link), $\omega_0\left(x\right)$, can be determined from \eqref{eq:Tackas_distribution} as 
\begin{align}
\omega_{0}\left(x\right) 
												 & = \sum_{n=0}^{\infty} {H}_{n} \left( x \right
) \!\!\int_{0}^{\infty} \!\!\!\Pr\{N^e\left(t-x\right) = n\}f_{t_z}(t) \textnormal{d}t \nonumber \\
												 & = \sum_{n=0}^{\infty} {H}_{n} \left( x \right) \Pr\{N_{t_z}^e \left( x \right) = n\}  
\label{eq:w_0}
\end{align}
where from \eqref{eq:H_n_R} and \eqref{eq:Nt}, $\Pr\{N^e\left(t-x\right) = n\} = G_n^{R}\left(t-x\right) - G^{R}_{n+1}
\left(t-x\right)$ is the PMF of number of renewals in an equilibrium renewal 
process over a fixed time and $\Pr\{N_{t_z}^e\left( x \right) = n\}$ is the same measure in a random time.

On the other hand, if $\chi\left(t_0\right)=1$, the collision-time 
distribution, $\omega_1\left(x\right)$, from \eqref{eq:Tackas_distribution} is
\begin{align}
\omega_{1}\left(x\right) 
%& \triangleq \int_{0}^{\infty} \Pr \left\{\beta\left(t\right)\leq x | \chi\left(t\right)=1\right\} f_{t_z}(t)\textnormal{d}t \nonumber \\
												 %&
												%\begin{aligned}
												 %= \sum_{n=0}^{\infty} {H}_{n+1}^{R} \left( x \right) \int_{0}^{\infty} & \big[G_n\left(t-x\right) - G_{n+1}\left(t-x\right)\big] \\ & \cdot f_T(t)dt
												%\end{aligned}
												%\nonumber \\
												 & = \sum_{n=0}^{\infty} {H}_{n}^{R} \left( x \right)\!\! \int_{0}^{\infty} \!\!\!\Pr\{N^o(t-x) = n\}f_{t_z}(t) \textnormal{d}t \nonumber \\
												 & = \sum_{n=0}^{\infty} {H}_{n}^{R} \left( x \right) \Pr\{N_{t_z}^o\left( x \right) = n\} 
\label{eq:w_1}
\end{align}
where $\Pr\{N^o(t-x) = n\} = G_n\left(t-x\right) - G_{n+1}\left(t-x\right)$ 
corresponds to the PMF of the number of renewals in an ordinary renewal process in a fixed time and $\Pr\{N_{t_z}^o \left( x \right)= n\}$ denotes the PMF in a random time.

As, at an arbitrary time instant $t_0>0$, we find the system in \emph{on} state 
with probability $\Pr\left\{\chi\left(t_0\right)=1\right\} = \frac{\bar{\eta}}{\bar{\eta} + \bar{\xi}} \triangleq \alpha$
and in \emph{off} state with probability $1-\alpha$,  the joint 
collision-time distribution function is 
\begin{equation}
\Omega\left(x\right) \triangleq \alpha\omega_1\left(x\right) + \left(1-\alpha\right)\omega_0\left(x\right) .
\label{eq:CT_Model}
\end{equation}

\section{Collision Time Analysis}
\label{sec:CollisionTimeAnalysis}

In this section, for a random packet length $t_z$ we 
find the \emph{on} time distributions, ${H}_{n} \left( x \right)$ and 
${H}_{n}^{R}\left( x \right)$, and the PMF of the number of renewals, 
$\Pr\{N_{t_z}^e \left(x\right)= n\}$ and $\Pr\{N_{t_z}^o \left(x\right)= n\}$, assuming various 
\emph{on}- and \emph{off}-time distributions of the WLAN system. 
 
\subsection{Interference On Time Distribution}
\label{sec:ON_time_distribution}

The packet transmission time, which depends on the packet length and the bit 
rate, corresponds to the \emph{on} time of an alternating renewal process.
In the following, we consider constant and exponentially distributed \emph{on}
 time without the loss of generality.  

\subsubsection{Constant packet length}
Assuming a constant WLAN packet length (i.e., $\bar{\eta} = t_w$), the \emph{on} 
time distribution is given by \cite{pcmAamir}
\begin{equation}
 {H}_{n} \left( x \right) = 
 \begin{cases}
 0, & x < nt_w\\
 1, & x \geq nt_w
 \end{cases}
\label{eq:H_n_x}
\end{equation} 
Since the residual time, $\eta^R$ is uniformly distributed in the interval 
$\left[0,t_w\right]$, we have \cite{pcmAamir}
\begin{equation}
 {H}_{n}^{R} \left( x \right) = 
 \begin{cases}
 0, & x < \left(n-1\right)t_w\\
 {\frac{{x-(n-1)T}}{{T}}}, & \left(n-1\right)T \leq x < nt_w\\
 1, & x \geq nt_w
 \end{cases} 
\label{eq:H_n_x_R}
\end{equation} 
%The constant packet length is assumed to simplify the analysis without the 
%loss of generality of the approach. For a given packet length distribution, 
%it is possible to obtain $n$-fold convolution and residual time distribution 
%of it.
\subsubsection{Random packet length}  
On the other hand, if $t_w$ follows the exponential distribution with 
parameter $\mu$ and mean $\bar{\eta} = 1/\mu$, 
${H}_{n} \left( x \right) = {H}_{n}^R \left( x \right)$ is the Erlang-$n$ 
distribution
\begin{equation}
{H}_{n} \left( x \right)= {H}_{n}^R \left( x \right) = 1 - \sum\limits_{k = 0}^{n-1} \frac{1}{k!} \left (\mu x \right )^k {\rm{exp}}\left (
-\mu x \right )
\label{eq:G_n_x_exp}
\end{equation}
 
\subsection{PMF of Number of Renewals}
\label{sec:OFF_time_distribution}

Let $P(t,z)$ be the probability generating function (PGF) of $N_t$, the number of renewals in a fixed interval in $(0,t)$, defined as
\begin{equation}
P(t,z) = \mathbf{E}\left[z^{N_t}\right] = \sum_{n = 0}^{\infty} \Pr\{N_t = n\} z^n 
\label{eq:P_tz}
\end{equation}
Then $P(z)$, the PGF of $N_{t_z}$ i.e., the number of renewals in a random interval in $(0,t_z)$ with PDF $f_{t_z}(t)$, is
\begin{equation}
P(z) = \int_0^\infty P(t,z)f_{t_z}(t) dt 
\label{eq:P_z}
\end{equation}
The PMF of $N_{t_z}$ from \eqref{eq:P_z} can be determined as
\begin{equation}
\Pr\{N_{t_z} = n\} = \frac{1}{n!} \frac{d^n}{dz^n} P\left(z\right) \Big\rvert_{z=0}, n = 0,1,2,...
\label{eq:N_tz}
\end{equation}

With the basic relations in place in \eqref{eq:P_tz}-\eqref{eq:N_tz}, we find $\Pr\{N_{t_z}^e\left( x \right) = n\}$ and $\Pr\{N_{t_z}^o\left( x \right) = n\}$ need in \eqref{eq:w_0} and \eqref{eq:w_1}. The PGF of the number of renewals in a fixed 
interval assumes a general form \cite[eq. (3.2.2)]{cox1970renewal}
\begin{equation}
P(t-x,z) = 1 + \sum_{n=1}^{\infty}{z^{n-1}(z-1) \{G_n (t-x)\}^n}
\label{eq:PGF}
\end{equation}
Now if the Laplace transform of $g_n(t)$ is $g^*_n(s)$, then that of $G_n(t-x)$ is $g^{*}_n(s)e^{-sx}/s$. The function $g^*_n(s)$ for ordinary and equilibrium renewal process is equal to $\{g^*(s)\}^{n}/s$ and to $\{1-g^*(s)\}\{g^*(s)\}^{n-1}/(\bar{\xi}s)$ respectively. Therefore, the Laplace transform of \eqref{eq:PGF} for equilibrium renewal process is 
\begin{equation}
P_e^*(s,z) = \frac{1}{s} + \frac{(z-1)\{1-g^*(s)\}}{\bar{\xi}s^2 \{1-z g^*(s)\}}e^{-sx}
\label{eq:P_e_s}
\end{equation}
while for ordinary renewal process we have 
\begin{equation}
P_o^*(s,z) = \frac{1-g^*(s)}{s\{1-z g^*(s)\}}e^{-sx}
\label{eq:P_o_s}
\end{equation}
 
Assuming $t_z$ is exponential distributed with parameter $\lambda_z$, from \cite[eq. (3.4.3)]{cox1970renewal} 
the \eqref{eq:P_e_s} and \eqref{eq:P_o_s} can easily be inverted with   
\begin{equation}
P_{\{e,o\}}(z) = \lambda_z P_{\{e,o\}}^*(s,z) \Big\rvert_{s=\lambda_z}
\label{eq:P_e_0_inv}
\end{equation}

Now by substituting \eqref{eq:P_e_s} and \eqref{eq:P_o_s} in \eqref{eq:P_e_0_inv}, one can find the PMFs of the number of renewals desired in \eqref{eq:w_0} and \eqref{eq:w_1} with \eqref{eq:N_tz}. We find that the PMF of number of renewals in 
\eqref{eq:w_0} is
%Now by substituting in \eqref{eq:P_e_0_inv} the expressions \eqref{eq:P_e_s} and \eqref{eq:P_o_s} for Laplace transform $P_e^*(s,z)$ and $P_o^*(s,z)$ for equilibrium and ordinary renewal processes, one can find the PMF of the number of renewals desired in \eqref{eq:w_0} and \eqref{eq:w_1} with \eqref{eq:N_tz}. We find that the PMF of number of renewals in \eqref{eq:w_0} is
%\begin{equation}
%P_e(z) = \lambda\left[\frac{1}{s} + \frac{(z-1)\{1-g^{*}(s)\}}{\bar{\xi}s^2\{1-zg^*(s)\}}e^{-sx}\right]_{s=\lambda}
%\end{equation} 
\begin{align}
& \Pr \{N_{t_z}^e \left(x\right) =n\} \nonumber \\ 
& \!\!\!\!= 
 \begin{dcases}
1 - \left[\frac{\{1-g^*(s)\}e^{-sx}}{s\bar{\xi}}\right]_{s=\lambda_z}, & \mkern-18mu \mkern-18mu \mkern-18mu \mkern-18mu n = 0 \\
	%\!\begin{aligned}
		 \left[\frac{e^{-sx}}{s\bar{\xi}}\left(\{g^*(s)\}^{n+1}\!\! - \!2\{g^*(s)\}^{n}\!\! +\!
															\{g^*(s)\}^{n-1}\right)\! \right]_{s=\lambda_z}\!\!\!\!, 
	%\end{aligned} 
	 \\ & \mkern-18mu \mkern-18mu \mkern-18mu \mkern-18mu n \geq 1
 \end{dcases}
\label{eq:N_e_n}
\end{align} 
while the PMF in \eqref{eq:w_1} is
%\begin{equation}
%P_o(z) = \lambda \left[\frac{1}{s} + \frac{(z-1)g^*(s)}{s \{1-zg^*(s)\} }e^{-sx}\right]_{s=\lambda}
%\end{equation} 
\begin{align}
& \Pr\{N_{t_z}^o \left(x\right)=n\} \nonumber \\
& = 
 \begin{dcases}
 1 - \left[{e^{-sx} g^*(s)}\right]_{s=\lambda_z}, & n = 0 \\
 \left[e^{-sx}\left(\{g^*(s)\}^n - \{g^*(s)\}^{n+1}\right)\right]_{s=\lambda_z}, & n \geq 1
 \end{dcases}
\label{eq:N_o_n}
\end{align} 

For any idle-time distribution of WLAN traffic with known Laplace transform, 
one can easily find the PMF of the number of renewals observed during a 
packet duration over the sensor link from \eqref{eq:N_e_n} and 
\eqref{eq:N_o_n}. In the following, we consider the exponential and hyperexponential 
idle-time distributions (without loss of generality) as examples.  
\subsubsection{Exponential idle times}
The Laplace transform of exponential distribution with parameter $\rho$ is $g^
{*}(s):=\frac{\rho}{s+\rho}$, and the distribution of the renewals is
\begin{align}
& \Pr\{N_{t_z}^e\left(x\right)=n\}= \Pr\{N_{t_z}^o\left(x\right)=n\} \nonumber
 \\
& = 
 \begin{dcases}
 1 - \frac{\rho e^{-\lambda_z x}}{\lambda_z + \rho}, & n = 0 \\
e^{-\lambda_z x} \left(\frac{\rho}{\lambda_z + \rho}\right)^{n-1}\left(\frac{
\lambda_z}{\lambda_z + \rho}\right), & n \geq 1
 \end{dcases}
\label{eq:N_o_n_e}
\end{align} 
\subsubsection{Hyperexponential idle time} The hyperexponential distribution 
is the mixture of $k$ exponential random variables, i.e.
\begin{equation} 
g(x) = \sum_{i=1}^{k}{p_i \rho_i e^{-\rho_i x}}
\end{equation}
where $\sum_{i=1}^{k} p_i = 1$ and $E[X] = \sum_{i=1}^{k} {p_i}/{\rho_i}$. 
The Laplace transform of hyperexponential PDF is,
\begin{equation} 
g^{*}(s) = \sum_{i=1}^{M} p_i\frac{\rho_i}{s+\rho_i}
\label{eq:hyperErlang_lap}
\end{equation}

Substituting \eqref{eq:hyperErlang_lap} into \eqref{eq:N_e_n} and 
\eqref{eq:N_o_n}, it is straightforward to find the desired expressions of PMFs, 
which are excluded here due to space limitations.

%\begin{equation}
%\Pr\left[N_T = n\right] = \int_0^\infty \Pr\left[N_t = n\right]f_T(t) dt 
%\end{equation}
%
%
%\begin{equation}
%f_T(t) = \lambda_s e^{-\lambda_s t}, t\geq 0
%\end{equation}
%
%\begin{equation}
%\Pr\left[N_T = n\right] = 1 - \frac{e^{-\lambda_s x}}{1+ \frac{\lambda_s}{\lambda}}
%\end{equation}
\subsection{Numerical Validation}

\begin{figure}[!t]
	\centering
		\includegraphics[width=1\linewidth]{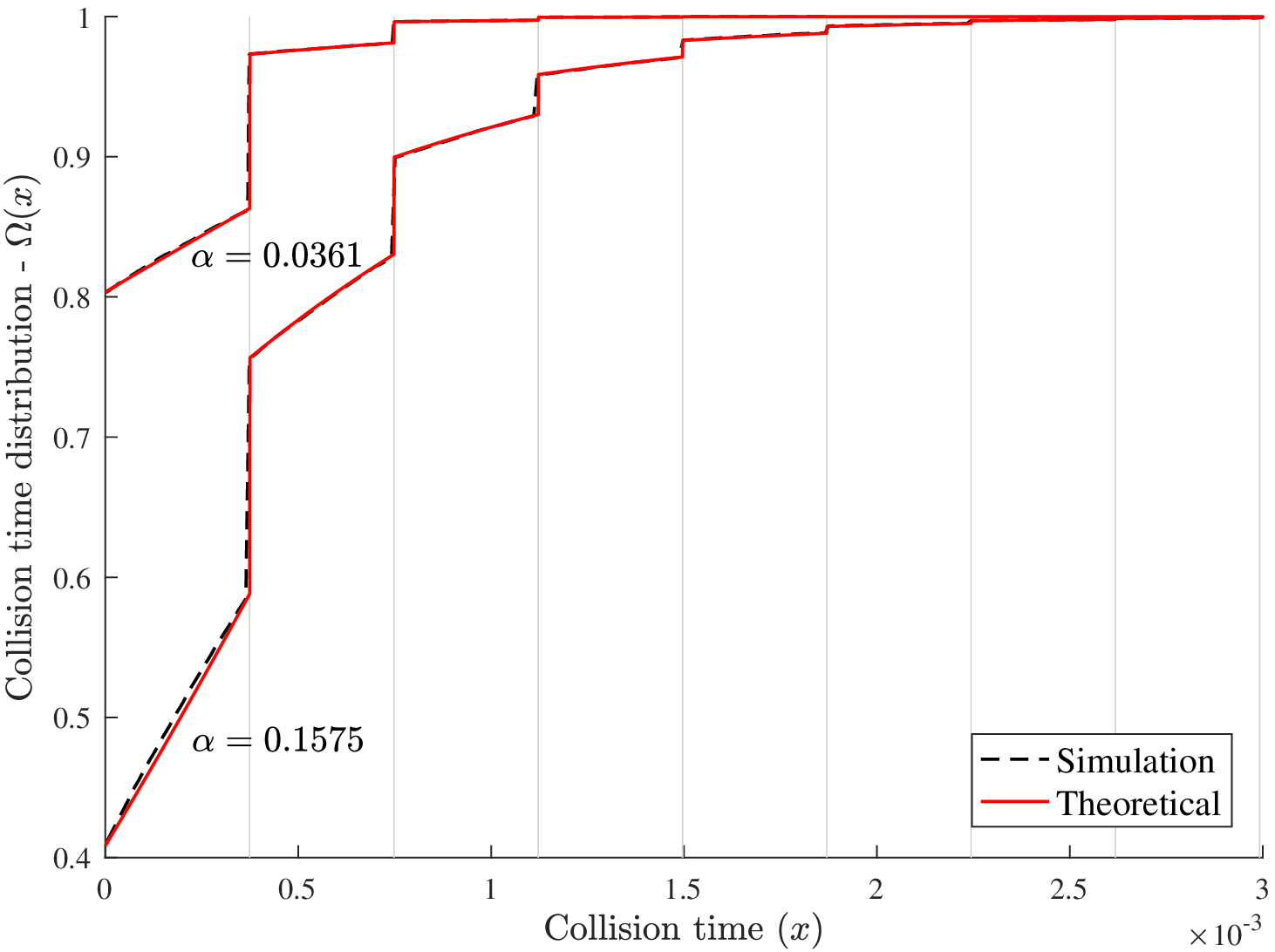}
	\caption{Collision-time distribution under exponential channel idle-times}
	\label{fig:Exponential_CTD}
\end{figure}

We validate the proposed CTD in \eqref{eq:CT_Model} with the distributions 
developed in Section ~\ref{sec:ON_time_distribution} \& 
\ref{sec:OFF_time_distribution} respectively. We assume that mean interfered 
packet 
transmission time is $\bar{t}_z=1/\lambda_z = 1. 984$ $m$s equaling a packet size of 
60 bytes at 256 kbps whereas WLAN busy-time is constant with $t_w=374$~$\mu$s 
which is equivalent of a nominal packet size of 500 bytes at 12 Mbps. Also, 
we assume that the idle-time is exponentially distributed. For validation, the numerical results from \eqref{eq:CT_Model}
 are compared against Matlab simulations, and shown in 
Fig.~\ref{fig:Exponential_CTD} for $\alpha = 0.0361$ and $\alpha = 
0.1575$. It can be observed that the numerical results 
are in excellent agreement with the simulations both for the low and high channel 
activity factors. Note that $y-$intercept is the probability of having no 
collisions with WLAN traffic during the interfered packet duration.

\begin{figure}[!t]
	\centering
		\includegraphics[width=1\linewidth]{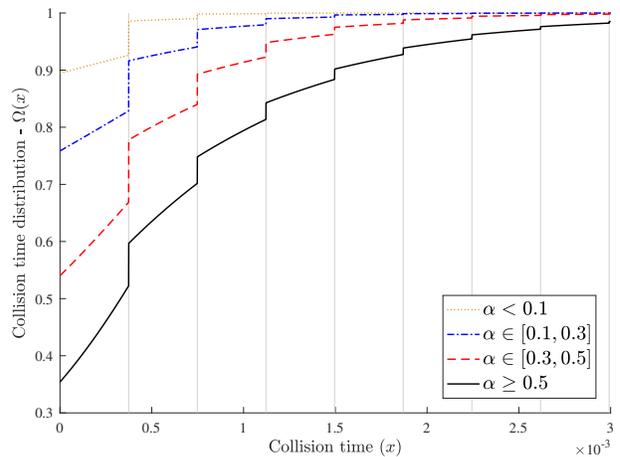}
	\caption{Collision-time distribution under hyperexponential channel idle-times}
	\label{fig:hyperExponential_CTD}
\end{figure}

After validating the proposed model, we look at the collision-time 
distributions under realistic hyperexponential idle-time distribution of WLAN. 
For this purpose, the hyperexpoential distribution parameters are taken 
from \cite{LucaStab} that fit best, based on the 
algorithm in \cite{feldmannFitting}, to the idle-time measurements taken from a real 
environment with heterogeneous WLANs/Bluetooth at 2.4 GHz. The fitted 
hyperexponential parameters with respect to the observed spectrum activity 
factor ($\alpha$) are given in Table~\ref{tab:table_A}. For the measurement setup and 
other details, the reader can refer to \cite{LucaStab}. Again, assuming the 
$\bar{t}_z=1/\lambda_z = 1. 984$ $m$s and $t_w=374$~$\mu$s, the numerical CTD is 
plotted in Fig.~\ref{fig:hyperExponential_CTD}. A number of observations can 
be made from Fig.~\ref{fig:hyperExponential_CTD}. As the activity factor 
increases the probability of collisions and the collision time increases. However owing 
to the long-tailed behavior in hyperexponential case, the probability of no collision remains smaller 
than the exponential case with the same activity factor. 

\begin{table}[!b]
\centering
 \caption{Estimated Parameters for Hyperexponentially Distributed Idle Times \cite{LucaStab}}
\label{tab:table_A}
\centering
\begin{tabular}{c|c|c|c|c}
 & $\alpha < 0.1$ & $\alpha \in [0.1, 0.3]$ & $\alpha \in [0.3, 0.5]$ & $\alpha \geq 0.5$\\
\hline
\hline
$1/\lambda_1$ & 0.040380 & 0.022490 & 0.012690 & 0.014890\\
$1/\lambda_2$ & 0.01174 & 0.006445 & 0.003289 & 0.002606\\
$1/\lambda_3$ & 0.00468 & 0.000388 & 0.000457  & 0.000395 \\
$p_1$   & 0.328        & 0.093 			 & 0.037 				& 0.012\\
$p_2$   & 0.356 				& 0.577 			 & 0.467        & 0.176\\
$p_3$   & 0.316 				& 0.330        & 0.496        & 0.812\\
\hline
\end{tabular}
\end {table}

\section{Packet Error Analysis under $\beta(t)$}
\label{sec:PER}

Depending on the composite traffic arrival process, $\beta(t)$ and the 
packet length of the sensor link, the number of interfered bits follow the 
collision-time distribution which we derived in the previous sections. In 
order to analyze the packet error performance under an observed $\beta(t)$, 
now we develop a packet error rate model (PER). 

\subsection{PER Model}

We assume that (in general) the downlink traffic from the WLAN AP to the 
stations outweighs the uplink traffic. In addition, when the WLAN stations are 
approximately at the same distance relative to the sensor system, it can be 
assumed that the interference power experienced by the sensor link is 
equivalent to the interference $I$ from the AP (see Fig.~\ref{fig:SysModel}). 
Furthermore, we consider the case where the desired signal undergoes constant 
channel gain as in \cite{hamidPDR}, while the WLAN interfering signal is 
subject to Rayleigh multipath fading. Therefore, without WLAN interference, the 
received signal-to-noise ratio (SNR) 
of the sensor link is $\gamma_s = E_s/N_0$ where $E_s$ is the 
bit energy of the desired signal and $N_0$ the noise power. Whereas in the presence 
of interference, the received SINR is $\gamma_{c} = {\gamma_s}/{(1+\gamma_I)}$
 \cite{INR} where $\gamma_I = |h_I|^2 E_I/N_o$ is the instantaneous 
interference-to-noise-ratio (INR) of the interfering signal with 
$\bar{\gamma}_I = E_I/N_o$ the average INR. Here, $h_I$ is the 
complex channel gain between the WLAN signal and the sensor receiver (with its 
envelop following the Rayleigh distribution). Therefore, $\gamma_I$ is 
exponentially distributed with the PDF, 
$f_{\gamma_I} = \frac{1}{\bar{\gamma}_I} \exp\left(-\gamma_I / \bar{\gamma
}_I\right)$.

\begin{figure*}[!t]
\normalsize
\begin{equation}
P_e(\gamma, \bar{\gamma}_I) =1 -\sum\limits_{\ell=0}^{N = \infty}\big(q_0\left(\gamma_s\right)\big)^{N-\ell} \int_0
^\infty \Big(q_1\left({\gamma_s}/{\gamma_I}\right)\Big)^{\ell} f_{\gamma_I}{d}
\gamma_I \Big(\Omega\left(\ell t_b\right)- \Omega\left(\left(\ell-1\right) t_b\right)\Big)
\label{eq:PER_Model2}
\end{equation}
\hrulefill
\end{figure*}

Consider the packet transmission time of the sensor link $t_z$ is 
exponentially distributed. If the $t_b$ is the physical layer bit duration, 
the number of bits in a packet are $N = [0, \infty$]. Let $b_e(\gamma)$ be the BER in AWGN channel 
which has general form for M-ASK, M-PAM, MSK, M-PSK and M-QAM modulations as  
\begin{equation} 	
b_{e}(\gamma) = c_mQ\left(\sqrt{k_m\gamma}\right) 
\label{eq:ber} 
\end{equation} 
where $c_m$ and $k_m$ are the modulation-specific constants, and $Q(\cdot)$ 
is the Gaussian $Q$-function. Define $q_0\left(\gamma_s\right) = 1 - b_e\left(
\gamma_s\right)$ be the bit 
success probability when there is no interference, and $q_1\left({\gamma_s}/{
\gamma_I}\right) = 1 - b_e\left({\gamma_s}/{(1+\gamma_I)}\right)$ be the bit 
success probability under interference. The PER of an $N$-bit packet 
with $\ell$ interfered number of bits is then given by
\begin{equation} 	
P_{e}(\gamma_s, \!\bar{\gamma}_I) \!=\! 1\! - \!\big(q_0\left(\gamma_s\right)\big)^{\!\!N-
\ell}\!\!\!\! \int_0 
^\infty \!\!\!\! \Big(\!q_1\!\left({\gamma_s}/{\gamma_I}\right)\!\Big)^{\ell}\!\! f_{\gamma_I}{d}
\gamma_I  
\label{eq:PER_Model1} 
\end{equation}

Under $\beta(t)$, $\ell$ will follow a distribution, and the PER model 
\eqref{eq:PER_Model1} can be reformulated as in \eqref{eq:PER_Model2}, where $
\Omega\left(x\right)$ is the collision time distribution defined 
in \eqref{eq:CT_Model} with  $x$ the 
collision time. Note that $\Omega\left(x\right)= 0$ for $x<0$. 
We assume that the level of interference at the sensor receiver is such that 
the effect of thermal noise on link performance can be ignored.

%\begin{equation}
%P_e(\gamma, \bar{\gamma}_I) =1 -\sum\limits_{\ell=0}^{N}q_{0}^{N-\ell}q_{1}^{\ell}\Big(\Omega\left(\ell T_b\right)- \Omega\left(\left(\ell-1\right) T_b\right)\Big)
%\label{eq:PER_Model2}
%\end{equation}

\subsection{PER Approximations}

In this section, we develop approximations for the integral in 
\eqref{eq:PER_Model2}, which using \eqref{eq:ber} is 

\begin{equation}
I_\ell = \int_0^\infty \Big(1 - c_m Q\Big(\sqrt{{k_m\gamma_s}/{\gamma_I}}\Big)\Big)^{\ell} f_{\gamma_I}{d}\gamma_I  
\label{eq:PER_integral}
\end{equation}

Due to the polynomial of degree $\ell$, the integral in 
\eqref{eq:PER_integral} is difficult to evaluate without using $Q$-function 
approximations. One can apply the binomial expansion to the term $(1-Q(x))^{
\ell}$ and use either $Q^N$ approximation \cite{QNApprox} or exponential 
function based bounds to $Q$-function \cite{expoApprox}\cite{expoImproved}. 
However, the approximation in \cite{expoApprox} is not accurate and 
approximation in \cite{expoImproved} is not integrable with respect to 
$\gamma_I$ for $\ell \geq 2$ in Rayleigh fading \cite{QNApprox}. Therefore, 
we used the $Q^N$ approximation \cite{QNApprox}

\begin{equation}
Q^\ell(x) \simeq \sum_{k_1, k_2,\cdots, k_{n_a}} {K_{\ell} C_{\ell} x^{f_m} e^
{\frac{-{\ell}x^2
}{2}}} 
\label{eq:QN_Approx}
\end{equation}
where the summation is carried over all sequences of non-negative integers 
$k_1+ \cdots + k_{n_a} = \ell$ and, $K_{\ell}$, $C_{\ell}$ and $f_m$ are 
defined after \cite[(4)-(6)]{QNApprox}. Note that, the accuracy of 
\eqref{eq:QN_Approx} depends on $n_a$ with $n_a = 8$ being the reasonable choice. 
With binomial expansion and using 
\eqref{eq:QN_Approx}, the integral \eqref{eq:PER_integral} is evaluated as in 
\eqref{eq:PDR_int_sol1}, where $\delta = \frac{1}{4}\left(2-f_m\right)$ and 
$K_n(\cdot,\cdot)$ is the modified Bessel function of the second kind.

\begin{figure}[!t]
	\centering
		\includegraphics[width=1\linewidth]{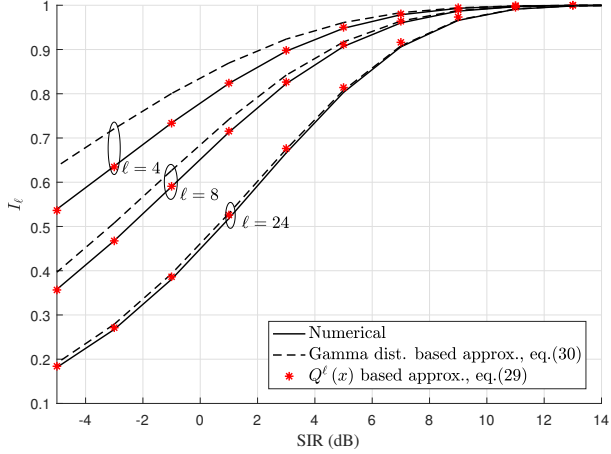}
		\vspace{-15pt}
	\caption{Comparing $I_{\ell}$ approximations for different values of $\ell$ with $c_m=1$ and $k_m=2$.}
	\label{fig:PERintegral}
\end{figure}

The approximation in \eqref{eq:PDR_int_sol1} is tight as shown in the 
Fig.~\ref{fig:PERintegral}, however, it is computationally intensive for higher integer powers 
($\ell \geq 8$) of the $Q$-function due to the fact that summation is carried 
over all sequences of non-negative integers $k_1, k_2, \cdots, k_{8}$ that 
sum to $\ell$. As a result, we utilized an extreme value theory based 
approximation proposed in \cite{perAamir} for higher powers. From 
\cite{perAamir}, the integrand in \eqref{eq:PER_integral}, denote as $I(x)$, can be as asymptotically 
approximated by the Gumbel distribution function for the sample maximum as
\begin{equation}
 I(x)  \simeq \mathrm{exp}\Big(\!\!-\mathrm{exp}\Big(-\frac{x-a_{\ell}}{b_{\ell}
}\Big)\Big)
\label{eq:gumbelDistributionApprox}
\end{equation}
where $a_{\ell} = \frac{2}{k_m} \big[\mathrm{erf}^{-1}\big(1 - \frac{2}{{\ell}
c_m}\big)\big]^2$ and $b_{\ell} = \frac{2}{k_m} \big[\mathrm{erf}^{-1}\big(1 - 
\frac{2}{{\ell}c_me}\big)\big]^2 - a_{\ell}$ are the normalizing constants, $e$ is the base 
of the natural logarithm and $\mathrm{erf}^{-1}(\cdot)$ is the inverse error 
function.

The approximation in \eqref{eq:gumbelDistributionApprox} is 
still not integrable in \eqref{eq:PER_integral}. However, Gumbel distribution 
function \eqref{eq:gumbelDistributionApprox} can be tightly approximated by 
the CDF of Gamma distribution by matching the first two moments:
\begin{equation}
\kappa_{\ell} = \frac{6\left(a_{\ell}+b_{\ell}E_0\right)^2}{\pi^2b_{\ell}^2}, 
\theta_{\ell} = \frac{a_{\ell}+b_
{\ell}E_0}{\kappa_{\ell}} \nonumber
\end{equation}
where $E_0 = 0.5772$ is the Euler constant. Using the CDF of Gamma 
distribution, the integral in \eqref{eq:PER_integral} becomes
\begin{equation}
I_\ell = \frac{1}{{\Gamma(\kappa_{\ell})}}\int_0^\infty {\acute{\gamma}\left(\kappa_{\ell},\frac{
\gamma_s}{\theta_{\ell}\gamma_I}\right)}
 f_{\gamma_I}{d}\gamma_I  
\label{eq:PER_final_gamma}
\end{equation}
%\begin{equation}
%\Big(1 - c_m Q\Big(\sqrt{{k_m\gamma_s}/{\gamma_I}}\Big)\Big)^{\ell} \approx 
%\frac{\acute{\gamma}\left(\kappa,\frac{
%\gamma}{\theta}\right)}{\Gamma(\kappa)}
%\label{eq:PER_final_gamma}
%\end{equation}
where $\acute{\gamma}\left(.,.\right)$ and $\Gamma(\cdot)$ stand for 
lower incomplete and complete Gamma functions respectively 
\cite[p.892]{table2007}. The integral in \eqref{eq:PER_final_gamma} over the exponential PDF, we get 
\eqref{eq:PER_Model_NEW_2}, where $K_n(\cdot,\cdot)$ is the modified Bessel 
function of the second kind.

\begin{figure*}[!t]
\normalsize
\begin{equation}
I_{\ell} = 1 + \frac{1}{\bar{\gamma}_I} \sum_{r = 1}^{\ell}{{\binom{\ell}{r}}\left(-c_m\right)^r \sum_{k_1, k_2,\cdots, k_{n_a}} K_r C_r 2^{1-\delta}
\Big(r k_m \gamma_s\bar{\gamma}_I\Big)^{\delta}\Big(k_m\gamma_s\Big)^{1-2\delta}K_n\left(-2\delta, \frac{\sqrt{2r k_m \gamma_s}}{\sqrt{\bar{\gamma}_I}}\right)}
\label{eq:PDR_int_sol1}
\vspace{-5pt}
\end{equation}
\vspace{-6pt}
\hrulefill
\end{figure*}
\begin{figure*}[!t]
\normalsize
\begin{equation}
I_{\ell} = \frac{1}{\Gamma\left(\kappa_{\ell}\right)} \left(\Gamma\left(\kappa_{\ell}\right) - 2 \left(\frac{\gamma_s}{\bar{\gamma}_I\theta_{\ell}}\right)^{\frac{\kappa_{\ell}}{2}} K_n\left(-\kappa_{\ell}, 2 \sqrt{\frac{\gamma_s}{\bar{\gamma}_I\theta_{\ell}}}\right)\right)
\label{eq:PER_Model_NEW_2}
\end{equation}
\vspace{-6pt}
\hrulefill
\end{figure*}

Fig.~\ref{fig:PERintegral} compares the proposed approximations of $I_{\ell}$ 
in \eqref{eq:PDR_int_sol1} and \eqref{eq:PER_Model_NEW_2} for different 
values of $\ell$. It can be observed that approximation 
\eqref{eq:PDR_int_sol1} is quite tight however its computational demanding. 
On the other hand, as the number of interfered 
bits ($\ell$) increase, the tightness of the approximation in  
\eqref{eq:PER_Model_NEW_2} also increases suggesting its usage for higher $
\ell$. 

\begin{figure}[!t]
	\centering
		\includegraphics[width=1\linewidth]{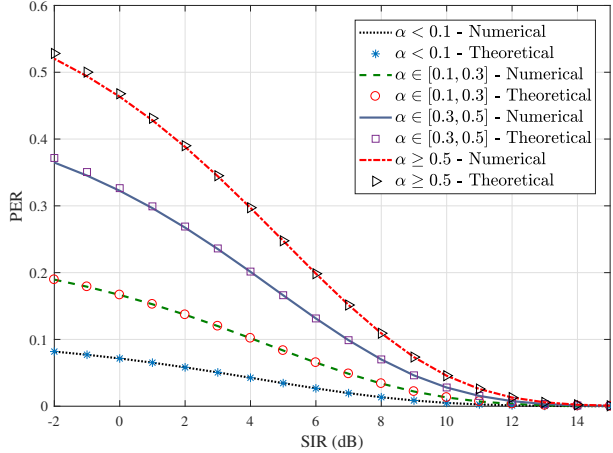}
		\vspace{-15pt}
	\caption{PER of an interfered link under WLAN interference. The modulation parameters of the interfered link are: $c_m=1$ and $k_m=2$, which correspond to BPSK/QPSK modulation.}
	\label{fig:PER}
	\vspace{-10pt}
\end{figure}

In Fig.~\ref{fig:PER}, the PER in \eqref{eq:PER_Model2} is evaluated under the 
collision-time distribution with hyperexpoential parameters given in Table~
\ref{tab:table_A}, and the approximations in \eqref{eq:PDR_int_sol1} and \eqref{eq:PER_Model_NEW_2}. 
As the approximation in \eqref{eq:PDR_int_sol1} is computationally demanding, 
we want to compute it for as lower values of $\ell$ as possible and use 
approximation \eqref{eq:PER_Model_NEW_2} for higher values. In 
Fig.~\ref{fig:PER}, for $\ell \leq 8$ we use \eqref{eq:PDR_int_sol1} and $\ell
> 8$ \eqref{eq:PER_Model_NEW_2}. It can be observed that this approach matches the 
numerical result tightly for small to large WLAN activity factors.      

\section{Conclusions}
\label{sec:Conclusions}

In this paper, we have developed a generic collision-time distribution (CTD) function of an 
interfered link based on renewal-theoretic modeling of the 
coexisting traffic. For the exponentially distributed interfered packet 
lengths, the CTD function is a random sum of the distributions of 
\emph{on-time} and \emph{number of renewals} of the coexisting traffic. The 
distribution of the number of renewals, which depends on the idle-time 
statistics, is derived theoretically. The distribution requires only the 
Laplace transform of the idle-time statistics thus can easily be evaluated 
for long-tailed hyperexponential or hyper-Erlang 
distribution. The theoretical collision-time distribution is in excellent 
agreement with the simulation results. We incorporated the proposed collision-
time distribution into a PER model which also takes into account the fading 
of the interfering signals. We investigated the approximations to PER and 
derived easy to compute and accurate expressions. As future extension of this 
work, we would like to study the potential of the packet collision model to 
RF energy harvesting problems.



%\bibliographystyle{IEEEtran}

\bibliographystyle{IEEEtran}
\bibliography{rtPCM}

\begin{thebibliography}{10}
\providecommand{\url}[1]{#1}
\csname url@samestyle\endcsname
\providecommand{\newblock}{\relax}
\providecommand{\bibinfo}[2]{#2}
\providecommand{\BIBentrySTDinterwordspacing}{\spaceskip=0pt\relax}
\providecommand{\BIBentryALTinterwordstretchfactor}{4}
\providecommand{\BIBentryALTinterwordspacing}{\spaceskip=\fontdimen2\font plus
\BIBentryALTinterwordstretchfactor\fontdimen3\font minus
  \fontdimen4\font\relax}
\providecommand{\BIBforeignlanguage}[2]{{%
\expandafter\ifx\csname l@#1\endcsname\relax
\typeout{** WARNING: IEEEtran.bst: No hyphenation pattern has been}%
\typeout{** loaded for the language `#1'. Using the pattern for}%
\typeout{** the default language instead.}%
\else
\language=\csname l@#1\endcsname
\fi
#2}}
\providecommand{\BIBdecl}{\relax}
\BIBdecl

\bibitem{ericsson2015}
A.~Ericsson, ``Ericsson mobility report: On the pulse of the networked
  society,'' \emph{Ericsson, Sweden, Tech. Rep. EAB-14}, vol. 61078, 2015.

\bibitem{LucaStab}
L.~Stabellini, ``Quantifying and modeling spectrum opportunities in a real
  wireless environment,'' in \emph{IEEE WCNC}.\hskip 1em plus 0.5em minus
  0.4em\relax IEEE, 2010, pp. 1--6.

\bibitem{geirhofer2006measurement}
S.~Geirhofer, L.~Tong, and B.~M. Sadler, ``A measurement-based model for
  dynamic spectrum access in {WLAN} channels,'' in \emph{IEEE MILCOM}.\hskip
  1em plus 0.5em minus 0.4em\relax IEEE, 2006, pp. 1--7.

\bibitem{Shin}
S.~Y. Shin, H.~S. Park, and W.~H. Kwon, ``Mutual interference analysis of
  {IEEE} 802.15.4 and {IEEE} 802.11b,'' \emph{Comput. Netw.}, vol.~51, no.~12,
  pp. 3338--3353, Aug. 2007.

\bibitem{pcmAamir}
A.~Mahmood, H.~Yigitler, and R.~J\"{a}ntti, ``Stochastic packet collision
  modeling in coexisting wireless networks for link quality evaluation,'' in
  \emph{IEEE ICC}, June 2013, pp. 1915--1920.

\bibitem{lee2013opportunistic}
S.~Lee, R.~Zhang, and K.~Huang, ``Opportunistic wireless energy harvesting in
  cognitive radio networks,'' \emph{IEEE Trans. W. Commun.}, vol.~12, no.~9,
  pp. 4788--4799, 2013.

\bibitem{Takacs1957}
L.~Tak{\'a}cs, ``On certain sojourn time problems in the theory of stochastic
  processes,'' \emph{Acta Mathematica Hungarica}, vol.~8, pp. 169--191, 1957.

\bibitem{cox1970renewal}
D.~Cox, \emph{Renewal Theory}.\hskip 1em plus 0.5em minus 0.4em\relax Methuen,
  1970.

\bibitem{feldmannFitting}
A.~Feldmann and W.~Whitt, ``Fitting mixtures of exponentials to long-tail
  distributions to analyze network performance models,'' \emph{Performance
  evaluation}, vol.~31, no. 3-4, pp. 245--279, 1998.

\bibitem{hamidPDR}
H.~Shariatmadari, A.~Mahmood, and R.~Jantti, ``Channel ranking based on packet
  delivery ratio estimation in wireless sensor networks,'' in \emph{IEEE WCNC},
  April 2013, pp. 59--64.

\bibitem{INR}
P.~S. Bithas and A.~A. Rontogiannis, ``Mobile communication systems in the
  presence of fading/shadowing, noise and interference,'' \emph{IEEE Trans.
  Commun.}, vol.~63, no.~3, pp. 724--737, March 2015.

\bibitem{QNApprox}
Y.~Isukapalli and B.~D. Rao, ``An analytically tractable approximation for the
  {G}aussian {Q}-function,'' \emph{IEEE Commun. Lett.}, vol.~12, no.~9, pp.
  669--671, September 2008.

\bibitem{expoApprox}
M.~Wu, X.~Lin, and P.~Y. Kam, ``New exponential lower bounds on the {G}aussian
  {Q}-function via {J}ensen's inequality,'' in \emph{IEEE 73rd Veh. Tech. Conf.
  (VTC Spring)}, May 2011, pp. 1--5.

\bibitem{expoImproved}
G.~K. Karagiannidis and A.~S. Lioumpas, ``An improved approximation for the
  {G}aussian {Q}-function,'' \emph{IEEE Commun. Lett.}, vol.~11, no.~8, pp.
  644--646, August 2007.

\bibitem{perAamir}
A.~Mahmood and R.~J\"{a}ntti, ``Packet error rate analysis of uncoded schemes
  in block-fading channels using extreme value theory,'' \emph{IEEE Commun.
  Lett.}, vol.~21, no.~1, pp. 208--211, Jan 2017.

\bibitem{table2007}
I.~S. Gradshteyn and I.~M. Ryzhik, \emph{Table of integrals, series, and
  products}, 7th~ed.\hskip 1em plus 0.5em minus 0.4em\relax Elsevier, 2007.

\end{thebibliography}

\end{document}